\documentclass[a4paper,fleqn,usenatbib]{mnras}

\usepackage[dvipdfmx]{graphicx}

\usepackage{amsmath}
\usepackage{color}

\usepackage[T1]{fontenc}
\usepackage{aecompl}
\usepackage{yfonts}

\newcommand{\lesssim}{\,\raisebox{-0.4ex}{$\stackrel{<}{\scriptstyle\sim}$}\,}
\newcommand{\gtrsim}{\,\raisebox{-0.4ex}{$\stackrel{>}{\scriptstyle\sim}$}\,}

\begin{document}

\title[]{A novel look at energy equipartition in globular clusters}

 \author[P. Bianchini et al.]{P. Bianchini$^{1,}$\thanks{E-mail:
bianchini@mpia.de}\thanks{Member of the International Max Planck Research School for Astronomy and Cosmic Physics at the University of Heidelberg, IMPRS-HD, Germany.},
G. van de Ven$^{1}$,
M. A. Norris$^{1,2}$,
E. Schinnerer$^{1}$,
\& A. L. Varri$^{3}$
\\
$^{1}$Max-Planck Institute for Astronomy, Koenigstuhl 17, 69117 Heidelberg, Germany\\
$^{2}$University of Central Lancashire, Preston, PR1 2HE, UK\\
$^{3}$School of Mathematics and Maxwell Institute of Mathematical Sciences, University of Edinburgh, King's Buildings, Edinburgh EH9 3JZ, UK
}

\date{Accepted 2016 March 2.}
\maketitle

\begin{abstract}
Two-body interactions play a major role in shaping the structural and dynamical properties of globular clusters (GCs) over their long-term evolution. In particular, GCs evolve toward a state of partial energy equipartition that induces a mass-dependence in their kinematics. By using a set of Monte Carlo cluster simulations evolved in quasi-isolation, we show that the stellar mass dependence of the velocity dispersion $\sigma(m)$ can be described by an exponential function $\sigma^2\propto \exp(-m/m_\mathrm{eq})$, with the parameter $m_\mathrm{eq}$ quantifying the degree of partial energy equipartition of the systems. This simple parametrization successfully captures the behaviour of the velocity dispersion at lower as well as higher stellar masses, that is, the regime where the system is expected to approach full equipartition. 
We find a tight correlation between the degree of equipartition reached by a GC and its dynamical state, indicating that clusters that are more than about 20 core relaxation times old, have reached a maximum degree of equipartition. This equipartition$-$dynamical state relation can be used as a tool to characterize the relaxation condition of a cluster with a kinematic measure of the $m_\mathrm{eq}$ parameter. Vice versa, the mass-dependence of the kinematics can be predicted knowing the relaxation time solely on the basis of photometric measurements. Moreover, any deviations from this tight relation could be used as a probe of a peculiar dynamical history of a cluster. Finally, our novel approach is important for the interpretation of state-of-the-art \textit{Hubble Space Telescope} proper motion data, for which the mass dependence of kinematics can now be measured, and for the application of modeling techniques which take into consideration multi-mass components and mass segregation.
\end{abstract}


\section{Introduction}

The apparent simplicity of galactic globular clusters (GCs) is the result of their  >10 Gyr evolution driven by the complex interplay of gravitational encounters (dynamical two-body interactions between stars), interactions with the host galaxy and internal stellar astrophysical processes. The deep understanding of these evolutionary ingredients is the key to interpret their current internal properties and to reveal their formation during the earliest epochs of galaxy formation.

In particular, gravitational encounters, over the two-body relaxation time-scale, are crucial in shaping the internal structural and dynamical properties of GCs. One of the effects of two-body interactions is to bring a system toward a state of thermalization, where stars with different masses approach the same energy \citep{Spitzer1987}. This is known as energy equipartition: massive stars lose kinematic energy sinking towards the center of the cluster, whilst, vice versa, low-mass stars gain kinetic energy and move toward the outer parts. This produces a mass-dependence of the kinematics with massive stars displaying a lower velocity dispersion than low-mass stars, and furthermore induces mass segregation.

Starting from the early work of \citet{Spitzer1969}, studies have been devoted to the understanding of how the process of energy equipartition takes place in GCs, pointing out that in a simple two-mass component system, energy equipartition is not always reached. Depending on the mass ratio of the particles of different species ($m_1/m_2$), as well as their contribution to the total mass of the system ($M_1/M_2$), the self-gravity of the heavier stars can dominate the potential in the core, and create a sub-system which is dynamically separated from the lighter components.  Such a sub-system will no longer be in energy equipartition with the rest of the system, and it may even undergo gravothermal collapse, while the light stars will not (``Spitzer instability''). This result was later generalized by  \citet{Vishniac1978} for the case of a continuous mass spectrum. 

Even in the Spitzer-stable case, many additional elements should be taken into account, in particular the fact that the distribution of the velocities of the stars in the system is affected also by the change of the potential itself, due to the change of the spatial distribution of the stars. Calculations have been performed with various models for the density-potential pairs, usually for the simple case of a two-component system (e.g., see \citealp{LightmanFall1978} the case of homogeneous spheres, or \citealp{InagakiWiyanto1984} for Fokker-Planck models). Relatively fewer studies have considered the evolution of multi-mass systems, but the lack of energy equipartition emerged very emphatically especially in the Fokker-Plank study by \citet{InagakiSaslaw1985}.  

This issue has been explored also by means of multi-mass N-body simulations, which have offered convincing evidence that collisional systems reach a state of only partial energy equipartition, especially in the outer regions (e.g., see \citealp{BaumgardtMakino2003}, Sect.~3.5; \citealp{Khalisi2007}). More recently, \citet{TrentivanderMarel2013} performed a systematic N-body study to characterize the dependence of the velocity dispersion on mass $\sigma(m)$, in terms of the scaling $\sigma \propto m^{-\eta}$ (where $\eta = 0.5$ corresponds to full equipartition). They find that $\eta< 0.5$, i.e. corresponding to only partial energy equipartition. Moreover, the lack of energy equipartition has also been tested with direct N-body simulations in the regime of open clusters \citep{Spera2016}. Finally, an additional confirmation of the lack of energy equipartition in globular clusters comes also from the side of distribution function-based models, especially lowered isothermal multi-mass equilibria (see Appendix).   

Even though GCs are not in full energy equipartition, the mass dependence of kinematics represents an additional complication to take into consideration for a complete understanding of the current internal dynamics of GCs. In fact, traditional modeling techniques that do not take into consideration this mass dependence present limitations that in general should be fully explored \citep{ShanahanGieles2015,Sollima2015}. Secondly, the evidence of mass-dependent kinematics should motivate the development and the application of multi-mass models, which could provide a more comprehensive and realistic description of the internal dynamics of GCs (e.g., the multi-mass generalization of the classic King models, proposed by \citealp{DaCostaFreeman1976} or the recently developed family of multi-mass lowered isothermal models by \citealp{GielesZocchi2015}).

Mass-dependent kinematics is now within reach of our observational capabilities, thanks to the combination of traditional spectroscopic-based line-of-sight velocities and high-precision \textit{Hubble Space Telescope} ($HST$) proper motions studies. In particular, the latter provide samples up to $100\times$ larger than the traditional line-of-sight velocity data sets, and allow us to measure the velocities for both giant stars and less-massive main sequence stars (see HSTPROMO data sets for 22 GCs, \citealp{Bellini2014, Watkins2015,Watkins2015b,Baldwin2016,Bianchini2016}; and references therein for other proper motion samples for specific GCs).

We therefore wish to introduce a novel approach for the analysis of energy equipartition in GCs suitable for applications to \textit{both simulations and observations}. Traditionally, the mass-dependent kinematics have been analyzed using the simple power-law dependence of the velocity dispersion on mass, $\sigma\propto m^{-\eta}$, that strictly is only valid for restricted stellar mass ranges \citep{TrentivanderMarel2013}. Fitting this function to simulations showed that the $\eta$ parameter is higher at the higher mass end (stellar remnants) than for the lower mass stars, indicating that a mass dependence of $\eta$ is, in fact, needed. Moreover, the analysis of the simulations has been limited to studies of clusters with fixed relaxation conditions \citep{TrentivanderMarel2013}, not allowing a direct comparison with real GC systems, characterized by a variety of relaxation conditions.

For this reason our work will be based on two premises. (1) The analysis of energy equipartition will be performed on a set of simulations all observed at a fixed time-snapshot. This gives us the advantage of creating a similarity to what we can actually observe, that is the Milky Way (MW) GCs that can be considered roughly coeval \citep{MeylanHeggie1997} and characterized by systems with a variety of relaxation states. (2) Extend the simple power-law $\sigma\propto m^{-\eta}$, introducing a more flexible function that can provide a fit to the mass-dependent velocity dispersion $\sigma(m)$ in the entire stellar mass range with a mass-dependent slope $\eta=\eta(m)$. Additionally, the function should provide a quantitative measure of the degree of energy equipartition reached by a system. The combination of the two points above will allow us to study the variety of mass-dependence of kinematics that we could expect for the MW GC system and to find possible correlations of the degree of partial equipartition with cluster properties.

In Sect. 2 we introduce the set of Monte Carlo cluster simulations used in this work and describe the construction of the $\sigma(m)$ profiles. In Sect. 3 the new fitting function is introduced and applied to the simulations. Section. 4 is devoted to the analysis of the results of the fits to the simulations and the study of how the degree of partial equipartition relates to cluster properties. In Sect. 5, we discuss the implication of our findings and, finally we summarize our conclusions in Sect. 6.


\begin{table}
\begin{center}
\caption{\textbf{Initial conditions of our set of simulations.} The original name of the simulations from \citet{Downing2010} are given in parentheses. We report the initial binary fraction $f_\mathrm{binary}$, the initial ratio of the intrinsic 3-dimensional tidal to half-mass radius $r_t/r_m$, the initial number of particles $N$, and the initial mass $M$. Simulations from \citet{Downing2010}, except Sim 7, 10low75-2M, from private communication of J. M. B. Downing.}
\tabcolsep=0.1cm

\begin{tabular}{lccccc}
\hline\hline
&$f_\mathrm{binary}$&$r_t/r_m$&N&M [$M_\odot$]\\
\hline

Sim 1 (10low75)&10\%&75&$5\times10^5$&$3.62\times10^5$\\
Sim 2 (50low75)&50\%&75&$5\times10^5$&$5.07\times10^5$\\
Sim 3 (10low37)&10\%&37&$5\times10^5$&$3.62\times10^5$\\
Sim 4 (50low37)&50\%&37&$5\times10^5$&$5.07\times10^5$\\
Sim 5 (10low180)&10\%&180&$5\times10^5$&$3.63\times10^5$\\
Sim 6 (50low180)&50\%&180&$5\times10^5$&$5.07\times10^5$\\
Sim 7 (10low75-2M)&10\%&75&$20\times10^5$&$7.26\times10^5$\\

\hline

\end{tabular}

\label{tab:0}
\end{center}
\end{table}

\begin{table*}
\tabcolsep=0.15cm
\begin{center}
\caption{\textbf{Projected properties of the set of simulations for the 4, 7 ,11 Gyr snapshots.} We report the concentration $c=\log(R_t/R_c)$, with $R_t$ and $R_c$ as projected tidal radius and projected core radius respectively, the half light radius $R_h$ in parsec, core radius $R_c$ in parsec, the logarithm of the half-light relaxation time  $T_\mathrm{rh}$ in yr, and the logarithm of the core relaxation time $T_\mathrm{rc}$ in yr. All simulations have an initial number of particles of N=500\,000, except for simulation 7 with N=2\,000\,000.}
\begin{tabular}{lrclrclrclrclrcl}
\hline\hline
 & \multicolumn{3}{c}{c} & \multicolumn{3}{c}{R$_h$} & \multicolumn{3}{c}{R$_c$} & \multicolumn{3}{c}{$\log T_\mathrm{rh}$} & \multicolumn{3}{c}{$\log T_\mathrm{rc}$}\\
&4 Gyr &7 Gyr &11 Gyr &4 Gyr &7 Gyr &11 Gyr &4 Gyr &7 Gyr &11 Gyr &4 Gyr &7 Gyr &11 Gyr &4 Gyr &7 Gyr &11 Gyr\\
\hline
Sim 1&1.52& 1.46& 1.45& 4.01&4.23&4.92& 2.74&3.12&3.15&9.382&9.487& 9.543&9.151&9.172&9.123\\
Sim  2&1.42&1.38 & 1.34& 4.89& 5.92&6.06& 3.42&3.62&3.89&9.474&9.579&9.655&9.345&9.287&9.286\\

Sim  3&1.26&1.21 & 1.16& 7.04& 8.16&9.05&4.92 & 5.52&6.07&9.658&9.755&9.820&9.647&9.656&9.645\\
Sim  4&1.21&1.16 & 1.12& 8.84&8.96&10.92&5.54 & 6.11&6.47&9.705&9.803&9.877&9.757&9.776&9.744\\

Sim  5&1.81& 1.95& 2.06& 1.53&1.90&2.69& 1.33&0.96&0.75&9.171&9.263&9.349&8.437&8.033&7.740\\
Sim  6&1.73&1.74 & 1.79& 2.96&3.10&3.05& 1.64&1.56&1.34&9.249&9.347&9.417&8.598&8.472&8.262\\

Sim 7 &1.52&1.52&1.51&2.57&2.62&2.90&1.73&1.87&1.85&9.415&9.498&9.565&9.040&8.965&8.991\\

\hline
\end{tabular}
\label{tab:1}
\end{center}
\end{table*}

\section{Simulations}
\label{sec:2}

We consider a set of Monte Carlo cluster simulations, developed by \citet{Downing2010} with the Monte Carlo code of \citet{Giersz1998} (see also \citealp{Hypki2013}). The simulations include an initial mass function, stellar evolution, primordial binaries, and a relatively high number of particles, providing a realistic description of the long-term evolution of GCs with a single stellar population.\footnote{Note that Monte Carlo simulations provide a high degree of realism achievable at low computational costs; moreover, they are consistent with direct N-body simulations \citep{Wang2016}.} No internal rotation is considered.

All simulations have their initial conditions drawn from a \citet{Plummer1911} model, a \citet{Kroupa2001} initial mass function, a metallicity of [Fe/H]=$-$1.3, and an initial tidal cut-off at 150 pc (making the simulations relatively isolated, comparable to halo clusters at $9-10$ kpc from the center of the MW). We consider a total of 6 simulations with 500\,000 initial particles, characterized by 3 values of initial concentrations and 2 values for the initial binary fraction (either 10\% or 50\%).  We also consider an additional simulation with 2\,000\,000 particles and 10\% initial binary fraction. All the snapshots that we will consider here are pre-core collapse\footnote{We restrict our investigations to pre-core collapsed systems since the interplay between mass segregation and core collapse is highly non-trivial; moreover, the majority of MW GCs are expected to be in a pre-core collapsed phase \citep{HarrisCat2010}.} and are indicative of typical metal poor GC spanning a large range of initial concentrations, binary fractions, and relatively high number of particles. The details of the initial conditions of the simulations are summarised in Table \ref{tab:0} and described in \citet{Downing2010}, expect sim 7 (10low75-2M), not present in the original work. The quantities used to characterized the initial conditions of the simulations are all intrinsic 3-dimensional quantities. The simulations were kindly shared by J. M. B. Downing.

We report in Table \ref{tab:1} the properties of the simulations typically assessed by observations, specifically for the time-snapshots at 4, 7, 11 Gyr. We report the concentration $c$ defined as $c=\log (R_t/R_c)$, with $R_t$ the projected tidal radius\footnote{Note that the projected tidal radius does not significantly differs from the 3-dimensional tidal radius, $r_t$.} and $R_c$ the projected core radius; the projected core radius $R_c$, defined as the radius where the surface density is half of the central surface density;\footnote{Calculated from number count surface density profiles.} the projected half-light radius $R_h$, containing half of the luminosity of the cluster; the logarithm of the half-mass relaxation time $T_\mathrm{rh}$ and the logarithm of the core relaxation time $T_\mathrm{rc}$. For the relaxation times, we follow the approach of the \citet{HarrisCat2010} catalog, that is Eq. (10) of \citet{Djorgovski1993} for the core relaxation time:
\begin{equation}
\label{eq:trc}
T_\mathrm{rc}=\frac{8.3377\times10^6\,\mathrm{yr}}{\ln(0.4\,N)} \left(\frac{M_\odot}{\langle m\rangle}\right) \left(\frac{\rho_0}{M_\odot/pc^3}\right)^{1/2} \left(\frac{R_c}{pc}\right)^3,
\end{equation}
with $N$ the number of stars in the cluster, $M$ the mass of the cluster, $\langle m\rangle$ the average stellar mass, and $\rho_0$ the central density of the cluster.\footnote{We define the central mass density of the cluster as the density enclosed within $R_h/10$.} For the half-mass relaxation time we use Eq. (8-72) of \citet{BinneyTremaine2008}:
\begin{equation}
T_\mathrm{rh}=\frac{6.5\times10^8\,\mathrm{yr}}{\ln(0.4\,N)} \left(\frac{M}{10^5M_\odot}\right)^{1/2} \left(\frac{M_\odot}{\langle m\rangle}\right) \left(\frac{R_h}{pc}\right)^{3/2}.
\end{equation}

\subsection{Construction of velocity dispersion-mass profiles}
\label{sec:profiles}
In order to quantify the mass dependence of the kinematics of our simulations, we construct the \textit{projected} velocity dispersion profile as a function of stellar mass, $\sigma(m)$. We restrict the analysis to stars within the projected half-light radius \footnote{GCs kinematics are typically observed within the half-light radius.} $R_h$ (considering a cylinder of radius $R_h$ around the z-axis as the line-of-sight direction) and include all the stars of our simulations within the mass range $0.1-1.8$ $M_\odot$ (the effect of different stellar objects such as binary stars and stellar remnants is separately discussed in Sect. \ref{sec:binaries})\footnote{The upper limit of 1.8 $M_\odot$ is chosen to guarantee a high enough number of stars per bin, since only a few stars have masses greater than that.}. For every 0.1 $M_\odot$ mass interval, we calculate the projected one-dimensional velocity dispersion and the associated errors, averaging the velocity dispersions of the three spatial coordinates. We point out that we consider projected quantities in order to enable a direct comparison with observations.

\begin{figure*}
\centering
\includegraphics[width=0.95\textwidth]{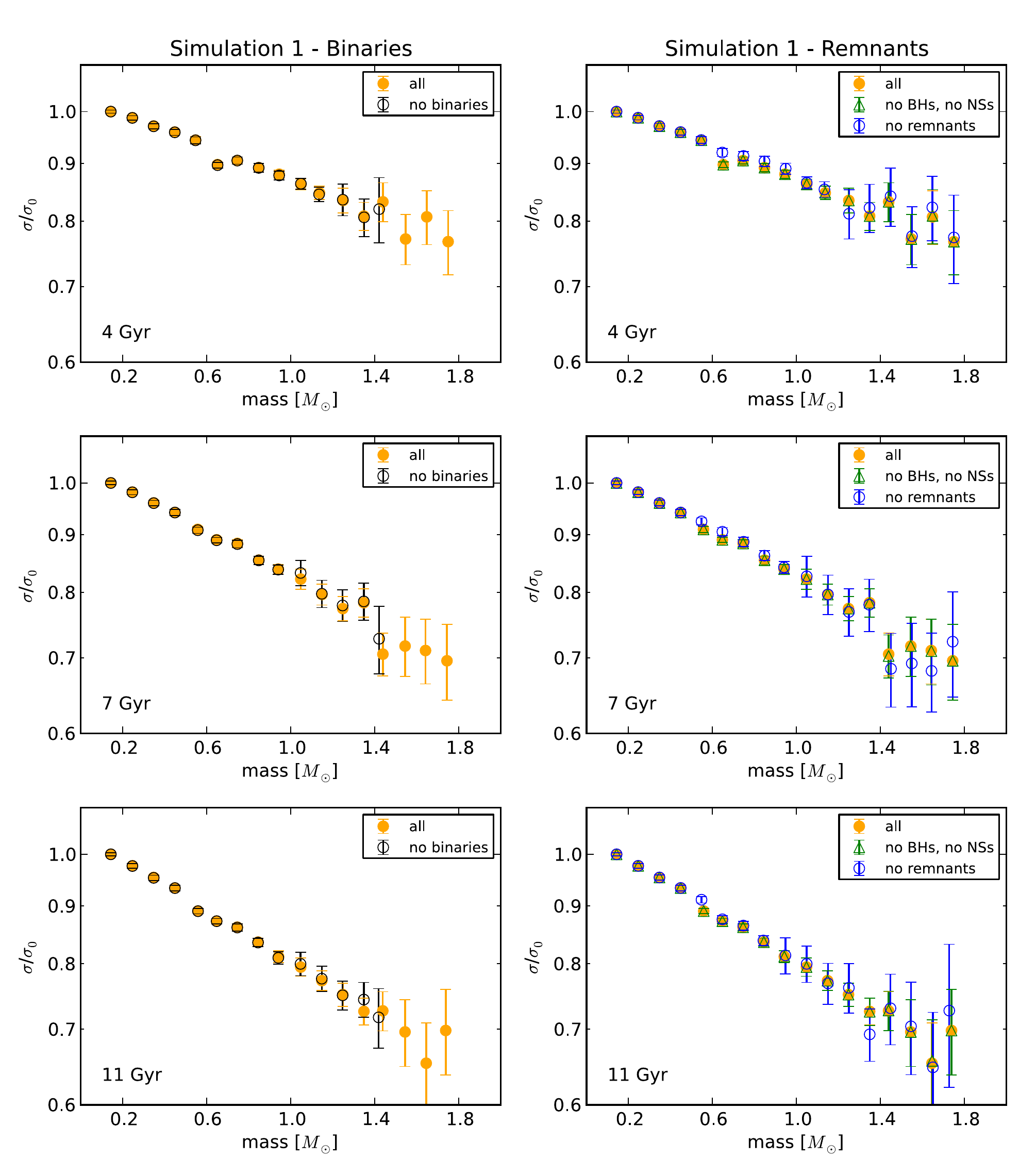}
\caption{Velocity dispersion as a function of stellar mass for the 4, 7, 11 Gyr snapshots of Simulation 1 (from top to bottom), restricted to the stars within the projected half-light radius. The profiles are normalized to the first bin, denoted as $\sigma_0$. \textbf{Left column:} Velocity dispersions for all stars within the half-light radius (orange circles) and when excluding binary stars only (open black circles). The binaries do not show an offset from the entire sample. \textbf{Right column:} Velocity dispersions for all stars within the half-light radius (orange circles), excluding dark stellar remnants (black holes and neutron stars; open green triangles), and excluding all stellar remnants (open clue circles). Stellar remnants do not introduce any significant bias in the velocity dispersions, except for white dwarfs around 0.6 $M_\odot$ (for details, see Fig. \ref{fig:2}).}
\label{fig:1}
\end{figure*}

\begin{figure*}
\centering
\includegraphics[width=0.95\textwidth]{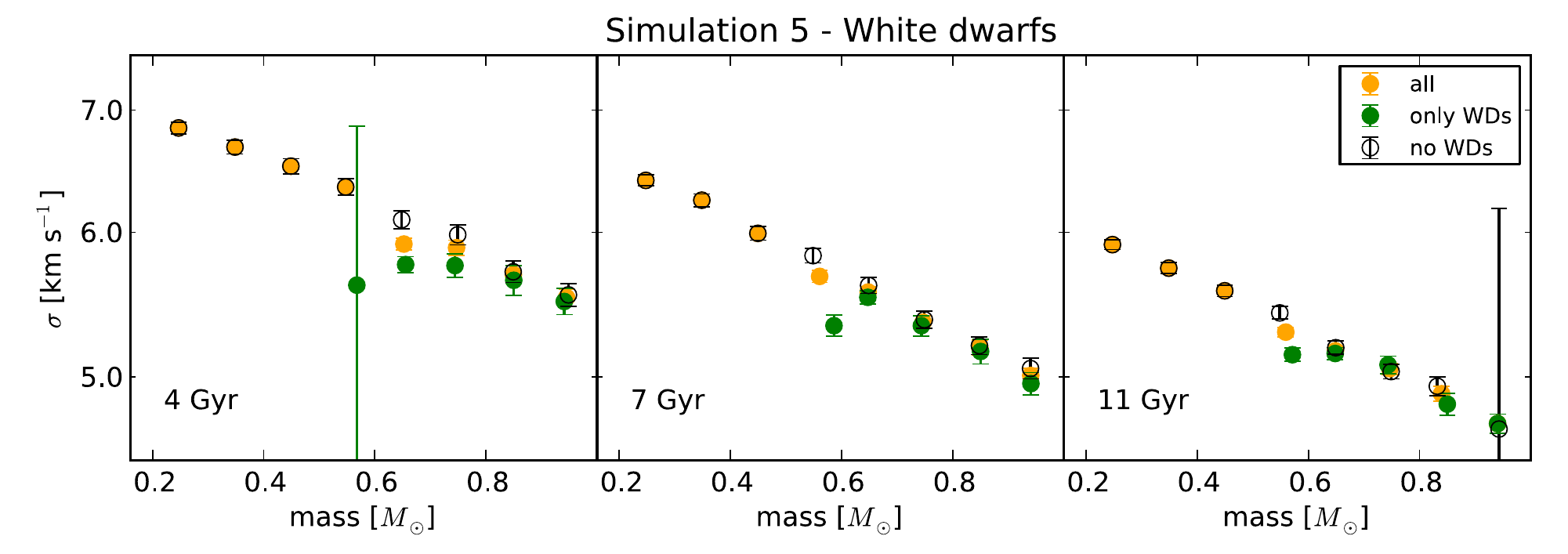}
\caption{Velocity dispersion as a function of stellar mass for the 4, 7, 11 Gyr snapshots of Simulation 5, for all stars within the half-light radius (orange circles), only white dwarf (green circles), and excluding white dwarfs (open black circles). The lowest-mass white dwarfs ($\approx0.6$ $M_\odot$) show a lower velocity dispersion than the other stars with similar mass, biasing the velocity dispersion of the sample with all stars towards lower values. The lower velocity dispersion of low-mass white dwarfs can be explained by the fact that they have not reached the same equipartition level as they recently underwent severe mass loss. The large error bars in the first and last panels are due to low number statistics.}
\label{fig:2}
\end{figure*}
\begin{figure*}
\centering
\includegraphics[width=0.95\textwidth]{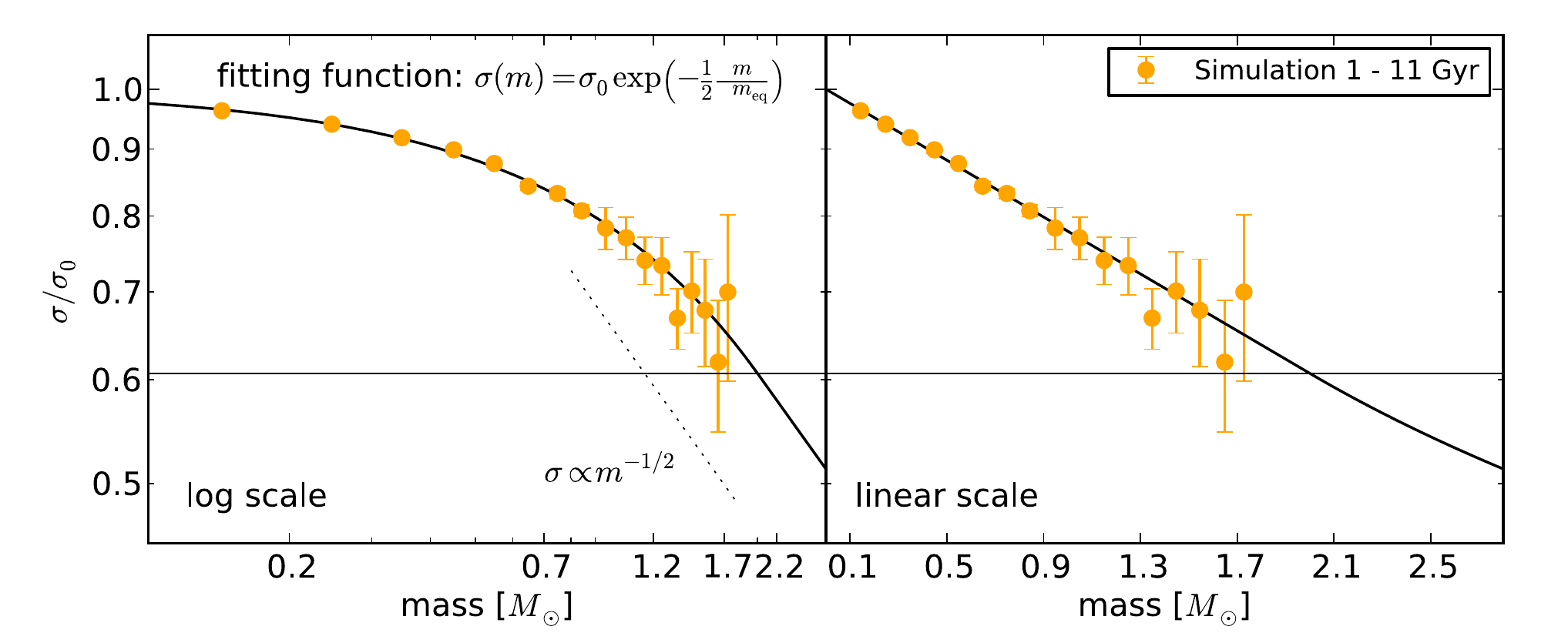}
\caption{Fit to the projected velocity dispersion as a function of the stellar mass $\sigma(m)$ using the exponential fitting function introduced with Eq. \ref{eq:1}. The free parameter $m_\mathrm{eq}$ indicates the degree of equipartition reached by the system. The fit is shown for Simulation 1 in both a log-log scale (left panel) and in a linear-log scale (right panel) and it is performed on all stars within the half-mass radius. The profiles are normalized at m=0, using the best fit parameter $\sigma_0$ (see Eq. 3). The dotted line shows the power-law function $\sigma\propto m^{-1/2}$ indicating full energy equipartition. The horizontal line intersects the fitting function at $m=m_\mathrm{eq}$; beyond this mass the slope of the fitting function is kept constant at $\sigma\propto m^{-1/2}$ (solid line). The exponential fitting function provides an excellent fit on all the sampled mass range.}
\label{fig:3}
\end{figure*}

\subsection{Role of binary stars and stellar remnants}
\label{sec:binaries}
In order to understand the role of different stellar objects in shaping the mass dependence of the velocity dispersion, we analyze separately the $\sigma(m)$ profile for binary stars and for stellar remnants. When considering binaries, we use the kinematics of their centre-of-mass.

Fig. \ref{fig:1} shows the result of our analysis: in the left panels the comparison between all stars and binary stars is reported, while in the right panels the comparison between the profiles with all stars, excluding dark remnants and excluding all remnants. From this we conclude that, for all the time-snapshots analyzed, binary stars and stellar remnants follow the same $\sigma(m)$ relation of single stars. However it is worth noting that around 0.6 $M_\odot$ the shape of $\sigma(m)$ shows a systematic dip. The right panels of Fig. \ref{fig:1} indicate that white dwarfs could be responsible for this feature, since the $\sigma(m)$ profile excluding all remnants does not show this dip around 0.6 $M_\odot$.

We investigate the effect of white dwarfs, by plotting in Fig. \ref{fig:2} separately all stars with and without white dwarfs and white dwarfs alone. White dwarfs that have recently formed and with masses around 0.6 $M_\odot$ underwent a severe mass loss. Their kinematics are not consistent with the one of other stellar objects with comparable mass, since they did not have time to dynamically relax. Therefore they are characterized by a lower velocity dispersion, in agreement with their original higher mass (see also \citealp{Heyl2015}). As the cluster evolves, two-body interactions slowly reduce this difference, as observed for the more evolved snapshots at 7 and 11 Gyr. 

In the following analysis we will construct $\sigma(m)$ profiles excluding white dwarfs, in order to avoid any bias.


\section{Fitting the velocity dispersion $-$ mass profile}
We wish to find a parametrization for the $\sigma(m)$ profile that describes the mass dependence of kinematics in the entire stellar mass range sampled and quantifies how close/far from energy equipartition the systems are. Traditionally a power-law $\sigma\propto m^{-\eta}$ has been used for limited ranges of masses, showing that the best fit parameter $\eta$ is higher for the higher mass end (stellar remnants) than for the lower mass stars \citep{TrentivanderMarel2013}. This indicates the differential behaviour of equipartition that is reached more efficiently in the higher-stellar mass regime ($\gtrsim1$ $M_\odot$) than in the lower-stellar mass regime, where the $\sigma(m)$ profile flattens ($\lesssim0.3$ $M_\odot$). It is therefore evident that a single $\eta$ value is not able to describe the entire trend of the $\sigma(m)$ profile and a fitting function with $\eta=\eta(m)$ is needed.

\subsection{Exponential fitting function}
We propose a simple exponential function, suitable for the entire stellar mass range sampled, and able to reproduce both the flat behaviour in the limit of low stellar masses and the steepening towards higher masses. A physical justification of the asymptotic limits is described in the Appendix. The function is characterized by a velocity scale parameter $\sigma_0$ and one mass scale parameter $m_\mathrm{eq}$:

\begin{equation}
 \sigma(m) = \left\{ \begin{array}{ll}
 \sigma_0\,\exp\left(-\frac{1}{2}\frac{m}{m_\mathrm{eq}}\right) & \mbox{if $m \leq m_\mathrm{eq}$,}\\
 \sigma_\mathrm{eq}\,\left(\frac{m}{m_\mathrm{eq}}\right)^{-1/2} & \mbox{if $m > m_\mathrm{eq}$.}\end{array} \right.
\label{eq:1}
\end{equation}
Here, $\sigma_0$ indicates the value of velocity dispersion at $m=0$, while $\sigma_\mathrm{eq}$ corresponds to the value of velocity dispersion at $m_\mathrm{eq}$, so that $\sigma_\mathrm{eq}=\sigma_0\exp\left(-\frac{1}{2}\right)$. The parameter $m_\mathrm{eq}$ quantifies the level of partial energy equipartition reached by the systems. For $m>m_\mathrm{eq}$ the system is characterized by constant full energy equipartition ($\sigma\propto m^{-1/2}$).

In accordance with the used power-law assumption ($\sigma\propto~m^{-\eta}$; \citealp{TrentivanderMarel2013}), the slope of our function is
\begin{equation}
\eta(m)=-\frac{d\ln\sigma}{d\ln m}=\left\{ \begin{array}{ll}
\frac{1}{2}\frac{m}{m_\mathrm{eq}} & \mbox{if $m \leq m_\mathrm{eq}$},\\
\frac{1}{2} & \mbox{if $m > m_\mathrm{eq}$.}\end{array} \right.
\label{eq:2}
\end{equation}
The truncation of the exponential function for $m>m_\mathrm{eq}$ was introduced in Eq. \ref{eq:1} in order to avoid values of the slope $\eta>1/2$ that would unphysically exceed energy equipartition as well as to match the asymptotic limits described in the Appedix, based on analytical multi-mass distribution function-based models \citep{GielesZocchi2015}.

The mass parameter $m_\mathrm{eq}$ will be used to quantify the degree of equipartition throughout our work: a system has reached equipartition in the stellar mass regime $m\gtrsim m_\mathrm{eq}$. Systems characterized by \textit{lower} values of $m_\mathrm{eq}$ are thus \textit{closer} to full energy equipartition.

\begin{figure}
\centering
\includegraphics[width=0.49\textwidth]{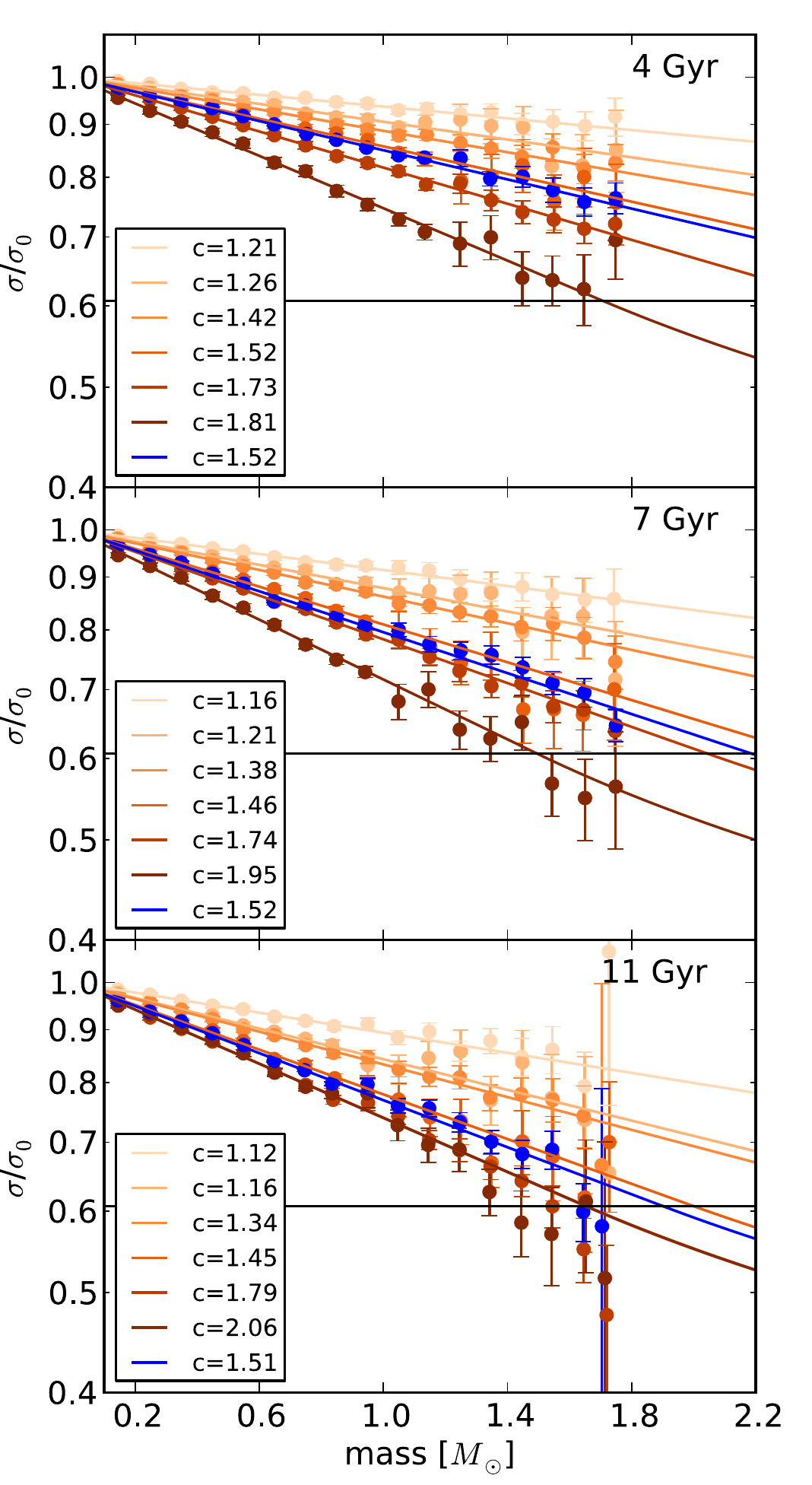}
\caption{Fit to the projected velocity dispersion as a function of stellar mass for the 4, 7, 11 Gyr snapshots of all our simulations. See Fig. \ref{fig:3} for details on the fit. The simulations are color coded according to their concentration (orange scale) with the exception of simulation 7 (blue dots), the only simulation with an initial number of particles of 2,000,000. The horizontal line intersects the fitting function at $m=m_\mathrm{eq}$. The simple exponential function fits well all our simulations in the entire mass range sampled. Minor deviations are observed exclusively in the high-mass regime of the most concentrated cluster model, which is likely about to reach the condition of core collapse (see bottom panel, c=2.06). Given a time snapshot, more concentrated clusters display a steeper velocity dispersion $-$ mass profile. Older snapshots have also steeper relation than younger ones, reflecting the dynamical evolution of the clusters. }
\label{fig:4}
\end{figure}

\begin{figure}
\centering
\includegraphics[width=0.4\textwidth]{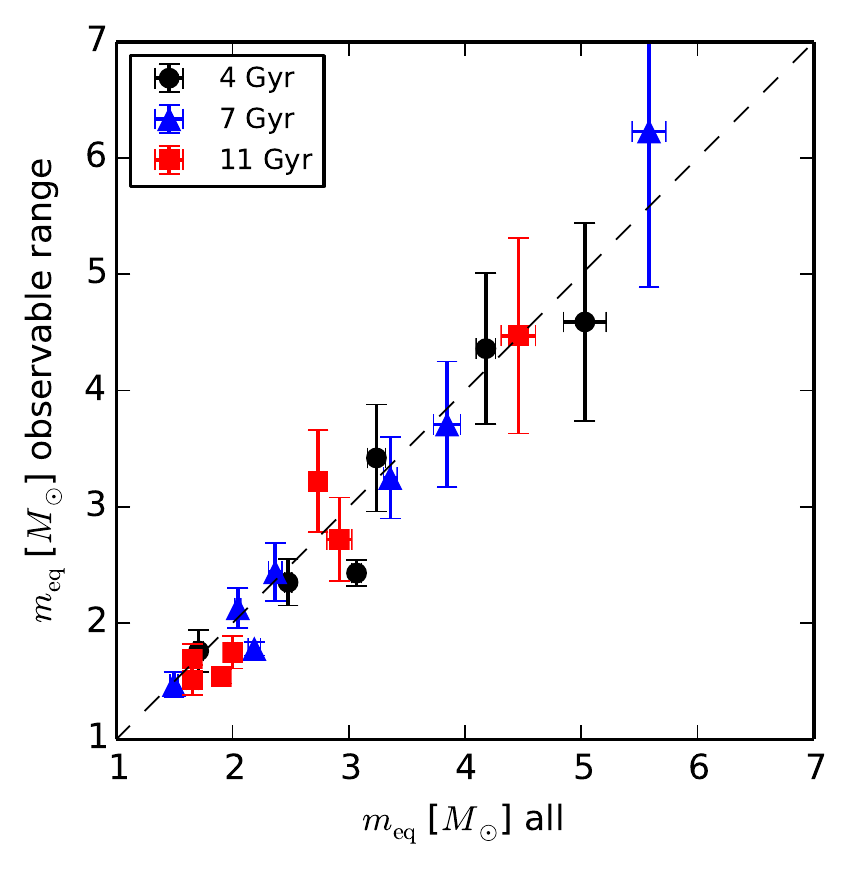}
\caption{Comparison of the $m_\mathrm{eq}$ parameter of our simulations obtained from the fits to the entire stellar mass range $0.1-1.8$ $M_\odot$ and from the fits (on discrete data, see Eq. \ref{eq:likelihood}) restricted to the observable mass range $0.4-1.0$ $M_\odot$. The fits to the observable mass range still allow for a good global description of the entire $\sigma(m)$ profile.}
\label{fig:5}
\end{figure}

\subsection{Application to the simulations}

We apply Eq. (\ref{eq:1}) to our set of simulations and quantify, through the parameter $m_\mathrm{eq}$, the degree of partial equipartition reached by the systems. We perform two fits: one using all stars in the mass range $0.1-1.8$ $M_\odot$ (excluding white dwarfs, as explained in Sect. \ref{sec:binaries}) and one restricting to only observable stars in the mass range $0.4-1.0$ $M_\odot$ (i.e., we exclude all stellar remnants) in order to match the typical observations.\footnote{Kinematic observations are now able to sample both bright (massive) stars and lower-mass stars along the main sequence. In particular, traditional spectroscopic line-of-sight measurements observe giant stars with masses $0.8-0.9$ $M_\odot$ (for $\sim10$ Gyr clusters) and proper motions provide the additional kinematic information for less massive main sequence stars, down to $\sim0.4$ $M_\odot$ (\citealp{Bellini2014,Watkins2015,Watkins2015b}); these could be complemented by the state-of-the-art line-of-sight velocities observations by MUSE@VLT, able to sample stars down to $\sim0.5$ $M_\odot$ \citep{Kamann2016}.}

The fit to all the stars is performed to the binned profiles (described in Sect. \ref{sec:profiles}) and will be used to demonstrate the performance of our fitting function. In the case restricted to observable stars only, we use a discrete fitting approach, which is particularly convenient and flexible for an application to real data, where errors or additional sources of contamination need to be included in the fit to the data. In this latter case, we define a likelihood function as
\begin{equation}
L_i=\prod_{i=1}^N \frac{1}{\sqrt{2\pi \sigma^2(m_i)}}\exp{\left[-\frac{1}{2} \frac{v^2_i}{\sigma^2(m_i)}\right]},
\label{eq:likelihood}
\end{equation}
where $\sigma(m_i)$ is given by Eq. (\ref{eq:1}), $m_i$ and $v_i$ are the stellar mass and the velocity of the observed stars, and the free parameters are $\sigma_0$ and $m_\mathrm{eq}$. 
Note that here we assume a Gaussian velocity distribution with zero mean velocity. The parameter space is explored using EMCEE, an affine-invariant Markov Chain Monte Carlo (MCMC) sampler \citep{Foreman-Mackey2013} and the mean and associated 1$-$sigma errors are returned. 

Fig. \ref{fig:3} shows the fit to the entire stellar mass range for one of the simulations. The left panel shows the $\sigma(m)$ profile in a log-log plot, while the right panel in a log-linear plot. The horizontal line intersects the $\sigma(m)$ profile at $m=m_\mathrm{eq}$, the mass above which the system is in equipartition. In Fig. \ref{fig:4}, we show the fits for all our simulations for the different time-snapshots and demonstrate that our fitting function provides an excellent description of the mass dependence of the kinematics in all cases. Minor deviations are observed exclusively in the high stellar mass regime of the most concentrated cluster model, which is likely about to reach the condition of core collapse (see bottom panel, dark orange line).

Table \ref{tab:2} summarizes the results of our fits to both the entire stellar mass range and to the restricted mass range. The two sets of fits give results consistent with each other, as also visualized in Fig. \ref{fig:5}. This indicates that using only a stellar mass range restricted to the current observations, it is still possible to obtain a good global description of the entire $\sigma(m)$ profile. In turn, this implies that we can also predict the mass dependence of the kinematics for both low-mass and high-mass stars for which the kinematics are not measurable (including non-observable dark remnants).

The values of $m_\mathrm{eq}$ obtained from the fits indicate that the systems are, as expected, only in partial energy equipartition, since the typical value of $m_\mathrm{eq}\gtrsim1.5$ $M_\odot$ indicates that all the stars sampled below this mass are characterized by a $\sigma(m)$ profile with a local slope $\eta<1/2$ (see Eq. \ref{eq:2}). In the following section we investigate in detail the relation between the degree of partial energy equipartition reached by a system and its global properties.

\begin{table}
\begin{center}
\caption{\textbf{Results of the exponential fit to the 4, 7, 11 Gyr snapshots of all our simulations.} The concentration c and the fitted parameters $\sigma_0$ and $m_\mathrm{eq}$ are reported with the associated 1-sigma errors. For every simulations two fits are performed: one for entire stellar mass range ($0.1-1.8$ $M_\odot$) and the other for only observable stars in a mass range similar to the one for which kinematic observations are available ($0.4-1.0$ $M_\odot$). The latter fit is performed to discrete data (see Eq. \ref{eq:likelihood}). Both fits give results consistent with each other.}
\tabcolsep=0.1cm
\begin{tabular}{lccccc}
\hline\hline
4 Gyr&&\multicolumn{2}{c}{all}&\multicolumn{2}{c}{observable}\\
&c&$\sigma_0$&$m_\mathrm{eq}$&$\sigma_0$&$m_\mathrm{eq}$\\
&&km s$^{-1}$&$M_\odot$&km s$^{-1}$&$M_\odot$\\

\hline
Sim 4 & 1.21&$4.50\pm0.01$ &$7.63\pm0.28$ & $4.48\pm0.04$&$7.36\pm1.37$ \\
Sim 3 & 1.26& $4.17\pm0.01$& $5.03\pm0.18$& $4.17\pm0.05$&$4.59\pm0.85$ \\

Sim 2 & 1.42& $5.82\pm0.01$& $4.18\pm0.08$&$5.77\pm0.07$ &$4.36\pm0.65$ \\
Sim 1 &1.52 & $5.46\pm0.01$& $3.24\pm0.08$&$5.41\pm0.07$ &$3.42\pm0.46$ \\
 
Sim 6 & 1.73&$7.28\pm0.01$ &$ 2.48\pm0.03$& $7.28\pm0.08$&$2.35\pm0.20$ \\
Sim 5 & 1.81& $7.38\pm0.03$& $1.71\pm0.04$& $7.30\pm0.13$&$1.76\pm0.18$ \\

Sim 7 & 1.52&$12.42\pm0.01$ &$3.07\pm0.04$ & $13.23\pm0.08$&$2.43\pm0.11$ \\

\hline

\end{tabular}
\vspace{0.2cm}

\begin{tabular}{lccccc}
\hline\hline
7 Gyr&&\multicolumn{2}{c}{all}&\multicolumn{2}{c}{observable}\\
&c&$\sigma_0$&$m_\mathrm{eq}$&$\sigma_0$&$m_\mathrm{eq}$\\
&&km s$^{-1}$&$M_\odot$&km s$^{-1}$&$M_\odot$\\

\hline
Sim 4 & 1.16& $4.22\pm0.01$&$5.58\pm0.14 $&$4.15\pm0.05$ &$6.23\pm1.34$ \\
Sim 3 & 1.21&$3.83\pm0.01 $&$3.84\pm0.11$ &$3.81\pm0.05$ &$3.71\pm0.54$ \\

Sim 2 & 1.38&$5.34\pm0.01$ &$3.36\pm0.06$ &$5.31\pm0.06$ &$3.25\pm0.35$ \\
Sim 1 & 1.46& $5.09\pm0.01$&$2.37\pm0.06 $&$5.05\pm0.07$ &$2.44\pm0.25$\\
 
Sim 6 & 1.74& $6.83\pm0.01$&$2.04\pm0.03 $&$6.74\pm0.08$ &$2.13\pm0.17$ \\
Sim 5 & 1.95&$ 6.94\pm0.03$& $1.50\pm0.03$& $6.95\pm0.12$&$1.47\pm0.11$ \\

Sim 7 & 1.52&$11.99\pm0.03$ &$2.19\pm0.05$ & $12.73\pm0.08$&$1.78\pm0.06$ \\

\hline

\end{tabular}
\vspace{0.2cm}

\begin{tabular}{lccccc}
\hline\hline
11 Gyr&&\multicolumn{2}{c}{all}&\multicolumn{2}{c}{observable}\\
&c&$\sigma_0$&$m_\mathrm{eq}$&$\sigma_0$&$m_\mathrm{eq}$\\
&&km s$^{-1}$&$M_\odot$&km s$^{-1}$&$M_\odot$\\

\hline
Sim 4 &1.12 &$ 3.89\pm0.01$&$4.46\pm0.15$ &$3.86\pm0.05$ &$4.47\pm0.84$ \\
Sim 3 & 1.16& $3.62\pm0.01$&$2.92\pm0.11$ &$3.63\pm0.05$ &$2.72\pm0.36$ \\

Sim 2 & 1.34&$5.12\pm0.01$ &$2.73\pm0.04 $&$4.99\pm0.06 $&$3.22\pm0.44$ \\
Sim 1 & 1.45& $4.88\pm0.01$ &$2.00\pm0.04 $&$4.97\pm0.07 $&$1.75\pm0.14$\\
 
Sim 6 & 1.79&$ 6.63\pm0.02$&$1.65\pm0.03$ &$6.55\pm0.10$ &$1.69\pm0.13$ \\
Sim 5 & 2.06&$6.34\pm0.03$ &$1.66\pm0.05$ &$6.43\pm0.12$ &$1.51\pm0.13$ \\

Sim 7 & 1.51&$11.42\pm0.02$ &$1.90\pm0.03$ &$12.20\pm0.09$ &$1.54\pm0.06$ \\

\hline

\end{tabular}

\label{tab:2}
\end{center}
\end{table}

\section{Degree of equipartition versus cluster properties}
\begin{figure}
\centering
\includegraphics[width=0.45\textwidth]{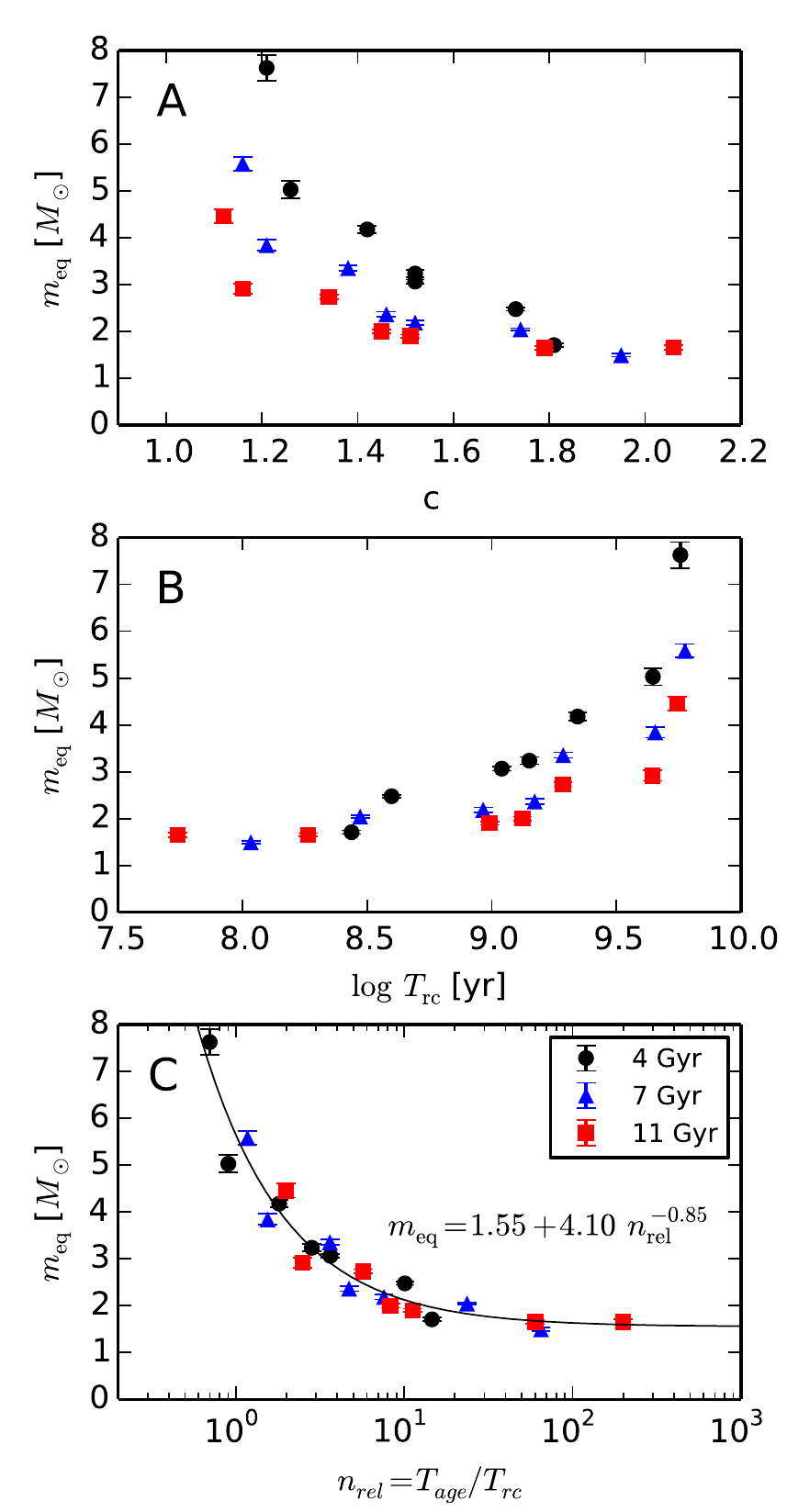}
\caption{Correlation between different cluster properties and the parameter $m_\mathrm{eq}$ obtained from the fits to the simulations. \textbf{From top to bottom:} the cluster concentration $c=\log(R_t/R_c)$ (panel A), the logarithm of the current core relaxation time $T_\mathrm{rc}$ (panel B), and the numbers of relaxation times experienced by a cluster $n_\mathrm{rel}=T_\mathrm{age}/T_\mathrm{rc}$, with $T_\mathrm{age}$ the age of the cluster (panel C). These plots demonstrate that the level of energy equipartition reached by a cluster depends on its relaxation condition. Well relaxed clusters (characterized by $n_\mathrm{rel}>20$) reach a maximum value for the degree of energy equipartition. The solid line in panel C is the best fit for the $m_\mathrm{eq}-n_\mathrm{rel}$ correlation, see Sect. \ref{sec:5}.}
\label{fig:6}
\end{figure}

The fitting function introduced in the previous section allows us to characterize the degree of partial equipartition reached by a cluster through the parameter $m_\mathrm{eq}$. We now analyze how this parameter correlates with structural properties of the GC simulations.

The first panel of Fig. \ref{fig:6} shows the relation between $m_\mathrm{eq}$ and the cluster concentration $c$, for the three time snapshots of 4, 7, 11 Gyr indicated by the different colours and symbols. More concentrated clusters are characterized by a lower value of $m_\mathrm{eq}$, corresponding to a steeper slope of the $\sigma(m)$ profile and, hence, are closer to energy equipartition (the increase of slope with concentration was already evident in Fig. \ref{fig:4}). The plot also shows that the equipartition-concentration relation depends on the age of the cluster: older clusters have reached a state closer to energy equipartition than younger clusters. This explains the three distinct relations, one for each time snapshot in the plot. Note, however, that in view of an application to MW GCs, any age dependence would only be marginally relevant, since all GCs can be safely considered as coeval.

In order to explain this relation between a purely photometric quantity (concentration) and a purely kinematic quantity (the degree of equipartition), we further investigate the role of the relaxation condition of the systems. For this reason, in the second panel of Fig. \ref{fig:6}, we plot $m_\mathrm{eq}$ against the current core relaxation time $T_\mathrm{rc}$ calculated from Eq. \ref{eq:trc}, for the particular time snapshot considered.\footnote{We use the core relaxation time since it provides a better description of the relaxation conditions for the central regions, in contrast with the half-light relaxation time that represents an average quantity suitable to describe the system globally.} Clusters with shorter relaxation times show lower values of $m_\mathrm{eq}$, indicating that two-body interactions have been more effective in establishing a higher degree of partial energy equipartition, since they have been acting for more relaxation times. However, a small dependence on the cluster age is still noticeable. 

In the bottom panel of Fig. \ref{fig:6}, we introduce the quantity $n_\mathrm{rel}=T_\mathrm{age}/T_\mathrm{rc}$, with $T_\mathrm{age}$ the age of the cluster. This quantity indicates the numbers of relaxation times that a cluster has experienced, with higher $n_\mathrm{rel}$ corresponding to more relaxed stellar systems.
A tight correlation is obtained for $m_\mathrm{eq}-n_\mathrm{rel}$, indicating clearly that the establishment of energy equipartition depends on the units of relaxation time experienced by a cluster. Interestingly, clusters with $n_\mathrm{rel}>20$ seem to reach an asymptotic maximum value of degree of equipartition, characterized by $m_\mathrm{eq}\simeq1.5$ $M_\odot$. These clusters have $\log T_\mathrm{rc}<8.5$, and are usually referred to as relaxed globular clusters, according to the classification of \citet{Zocchi2012}. A further discussion of the implications of the $m_\mathrm{eq}-n_\mathrm{rel}$ is presented in Sect. \ref{sec:5}.

\section{The dynamical state of a cluster}
\label{sec:5}
We have shown that the degree of partial equipartition primarily correlates with the relaxation condition of the clusters (number of relaxation times that a cluster has experienced). Additionally, a dependence on concentration and age is also visible.

The cluster simulations used in this work are all initialized with the same tidal cut-off that sets them into a relatively isolated initial condition, suitable for halo clusters at 9-10 kpc from the Galactic Center (see \citealp{Downing2010} for details). Moreover, note that the simulations did not undergo core collapse. Our work therefore does not take into consideration such a particularly complex phase of star clusters dynamical evolution. However, the homogeneity of our set of simulations still allows us to investigate the fundamental effects that are solely connected to the internal dynamical processes. 

Here, we focus primarily on the implications of the correlation between the relaxation condition of the cluster and the degree of energy equipartition. In fact, this is a more sound relation between two well connected internal dynamical properties of the clusters. It provides also a straightforward interpretation: more relaxed clusters have reached a higher degree of partial energy equipartition. 

This relation can provide a fundamental tool to measure the relaxation condition of a cluster. Relaxation time, as used in this work, is a quantity accessible observationally from solely photometric quantities and it is already available for MW GCs (\citealp{HarrisCat2010}). The degree of partial energy equipartition can be efficiently determined using the parameter $m_\mathrm{eq}$ of the fitting function defined in Eq. \ref{eq:1}, applied to a combination of line-of-sight velocities and state-of-the-art \textit{HST} proper motion data sets.

We fit a power-law to the $m_\mathrm{eq}-n_\mathrm{rel}$ relation in the bottom panel of Fig. \ref{fig:6} and obtain
\begin{equation}
m_\mathrm{eq}=(1.55\pm0.23)+(4.10\pm0.31)\,n_\mathrm{rel}^{-0.85\pm0.12},
\end{equation}
and the corresponding inverse function
\begin{equation}
n_\mathrm{rel}=5.28\pm1.35\,(m_\mathrm{eq}-1.55\pm0.23)^{-1.18\pm0.17}.
\end{equation}

Given the $m_\mathrm{eq}-n_\mathrm{rel}$ relation, with a measure of $n_\mathrm{eq}=T_\mathrm{age}/T_\mathrm{rc}$ is possible to predict the $m_\mathrm{eq}$ parameter, hence the mass dependence of the kinematics for a given globular cluster. In this way, the $m_\mathrm{eq}-n_\mathrm{rel}$ relation can be used to predict the dynamics of dark stellar remnants or of other stars for which the kinematics are not easily available (see \citealp{Baldwin2016} for an application to blue straggler stars, and \citealp{Bianchini2016} for binary stars). This is fundamental since it can allow to carry out a complete dynamical analysis without neglecting the effect of partial energy equipartition and mass dependent kinematics. In addition, it offers a reference framework for a direct comparison with multi-mass modeling techniques \citep[e.g.,][]{GielesZocchi2015}. Interestingly, \citet{GielesZocchi2015} also find a correlation between the degree of energy equipartition and concentration (defined as the central depth of the potential) for their recently developed multi-mass distribution function models (see first panel Fig.\ref{fig:6} and Appendix). 

Inversely, with a kinematic measure of $m_\mathrm{eq}$, one can predict the $n_\mathrm{rel}$ for a given cluster and therefore characterize its relaxation condition and provide a dynamical age, indicating at which stage of evolution the system is. Therefore, the correlation between energy equipartition and relaxation time found here offers an additional dynamical age estimator to be added to those introduced in the literature. In particular, \citet{Ferraro2012} introduced a dynamical clock calibrated on the radial distribution of blue stragglers stars. This is based on the fact that the radial distribution of blue stragglers stars is determined by mass segregation (as a consequence of two-body relaxation leading to partial energy equipartition) and therefore it depends on the dynamical age of the cluster. Our approach, purely based on kinematics, allows one to look at the same problem from an independent and complimentary perspective.

Finally, the relations shown in this work can potentially be used as a tool to highlight the complexity of the evolutionary path of a cluster. In fact, any deviations from the tight $m_\mathrm{eq}-n_\mathrm{rel}$ relation determined here for the non-rotating quasi-isolated clusters, could be used to infer a complex dynamical evolution of a particular cluster, highlighting for example post-core collapse clusters, presence of intermediate-mass black holes, clusters with a peculiar orbit around the MW, accretion vs. in-situ formation, or other peculiar formation environments (e.g., nucleus of dwarf galaxies). In parallel, also the relation between degree of equipartition and concentration (first panel Fig. \ref{fig:6}) can be used to single out complex evolutionary paths.


\section{Conclusions}
Two-body interactions shape the internal structure and dynamics of globular clusters over their long-term evolution, bringing the systems in a state of partial energy equipartition characterized by mass-dependent kinematics. In this work, we introduced a novel approach to characterize the degree of partial energy equipartition reached by GCs suitable for both simulations and observations, and we investigated its correlation with GCs properties.

We analyzed a set of Monte Carlo cluster simulations spanning a large range of concentrations and binary fractions, and considered them at the \textit{same time-snapshots} (in line with the MW GC system which is characterized by coeval clusters). For these simulations, we constructed the projected $\sigma(m)$ profile, describing the mass-dependence of the velocity dispersion, in the region within the half-light radius. We summarize our findings here.

\begin{itemize}
\item Different stellar objects (single stars, binary stars, stellar remnants) follow the same $\sigma(m)$ profile, with the exception of recently formed white dwarfs that underwent rapid severe mass loss. These white dwarfs are characterized by lower velocity dispersion than the one expected for their current mass.

\item \textit{Fitting function for mass-dependent kinematics.} We introduce a simple exponential fitting function able to match the mass dependence of the velocity dispersion in the entire stellar mass range sampled. This function is able to reproduce the flattening of the slope of the $\sigma(m)$ profile towards low stellar masses and the steepening at the higher-mass regime. The fitted parameter is the mass parameter $m_\mathrm{eq}$ that is physically well motivated as it indicates the degree of partial equipartition reached by the cluster. For $m\geq m_\mathrm{eq}$ the slope of the function corresponds to that of full energy equipartition; while for $m<m_\mathrm{eq}$ only partial equipartition is achieved. The exponential fitting function provides excellent fits to the mass-dependent kinematics of our simulations, showing that the systems are only in partial energy equipartition. Note that our function can be considered an extension of the commonly used power-law function $\sigma\propto m^{-\eta}$ that is instead only valid for restricted mass ranges.  

\item \textit{Applicability to observations.} We tested our fitting function on different mass ranges. Using a discrete fitting technique, we showed that, even for the restricted mass range $0.4-1.0$ $M_\odot$ accessible from state-of-the-art observations from combination of \textit{HST} proper motions and line-of-sight velocities, it is still possible to reliably characterize the degree of partial equipartition with the parameter $m_\mathrm{eq}$. This indicates that, once the $\sigma(m)$ profile is characterized, it can be used to predict the mass dependence of kinematics also for the non-observable low-mass regime and stellar remnant mass regime (see \citealp{Baldwin2016} for an application to blue straggler stars). This will be particularly useful to carry out comprehensive dynamical modeling for those clusters for which the kinematics is restricted to only a limited stellar mass range.

\item \textit{Measuring the dynamical state of a cluster.} We looked for correlations of the degree of energy equipartition given by the parameter $m_\mathrm{eq}$ with different cluster properties. In particular, we find that more concentrated clusters are closer to energy equipartition than less concentrated ones and that younger clusters are further away from energy equipartition than older ones. We showed that these relations are due to the correlation of the degree of energy equipartition with the relaxation state of the cluster, which we quantified by $n_\mathrm{eq}=T_\mathrm{age}/T_\mathrm{rc}$ with $T_\mathrm{age}$ the age of the cluster and $T_\mathrm{rc}$ the current core relaxation time. The tight relation obtained for $m_\mathrm{eq}-n_\mathrm{rel}$ can serve as a tool to investigate the dynamical condition of a GC. In fact, given a relaxation state of a cluster (easily accessible from photometric quantities), it is possible to predict the $m_\mathrm{eq}$ parameter, and therefore the mass-dependence of the kinematics. Vice versa, measuring the equipartition parameter $m_\mathrm{eq}$ from kinematics, it is possible to retrieve the internal dynamical state of a cluster. Finally, the validity of this relation is restricted to quasi-isolated clusters, so that any deviations from it could potentially be used as a simple tool to unveil a peculiar dynamical history of a given cluster (e.g., post-core collapse, presence of intermediate-mass black hole, disk shocking, in situ vs. accreted formation, peculiar formation environments). We plan to undertake a specific analysis in order to quantify these effects in a follow up work.

\end{itemize}

A natural consequence of energy equipartition is the sinking of massive stars into the central regions of a clusters (mass segregation). This causes a radial variation of the mass-to-light ratio, M/L, in a cluster. We therefore anticipate a dependence of M/L on the dynamical state of a cluster and hence on its degree of energy equipartition. We will address this point further in a forthcoming paper.
Finally, we point out that the approach introduced in this work to describe energy equipartition in GCs can serve as an optimal tool not only to characterize simulations and state-of-the-art kinematic observations, but also for testing dynamical models in which multi-mass components (i.e., a realistic mass function) are taken into consideration (e.g. \citealp{GielesZocchi2015}).

\section*{Acknowledgments}
We are grateful to Jonathan M. B. Downing for providing the Monte Carlo simulations used in this work. We wish to thank Giuseppe Bertin, Roeland van der Marel, Anna Sippel and Laura Watkins for useful comments and discussions. ALV is grateful to Douglas Heggie for many interesting discussions, and acknowledges financial support from the 1851 Royal Commission. We thank the referee for helping improving the clarity of our paper. This work was supported by Sonderforschungsbereich SFB 881 "The Milky Way System" (subproject A7 and A8) of the German Research Foundation (DFG).

\bibliographystyle{mnras} 
\bibliography{biblio} 

\begin{thebibliography}{}
\makeatletter
\relax
\def\mn@urlcharsother{\let\do\@makeother \do\$\do\&\do\#\do\^\do\_\do\%\do\~}
\def\mn@doi{\begingroup\mn@urlcharsother \@ifnextchar [ {\mn@doi@}
  {\mn@doi@[]}}
\def\mn@doi@[#1]#2{\def\@tempa{#1}\ifx\@tempa\@empty \href
  {http://dx.doi.org/#2} {doi:#2}\else \href {http://dx.doi.org/#2} {#1}\fi
  \endgroup}
\def\mn@eprint#1#2{\mn@eprint@#1:#2::\@nil}
\def\mn@eprint@arXiv#1{\href {http://arxiv.org/abs/#1} {{\tt arXiv:#1}}}
\def\mn@eprint@dblp#1{\href {http://dblp.uni-trier.de/rec/bibtex/#1.xml}
  {dblp:#1}}
\def\mn@eprint@#1:#2:#3:#4\@nil{\def\@tempa {#1}\def\@tempb {#2}\def\@tempc
  {#3}\ifx \@tempc \@empty \let \@tempc \@tempb \let \@tempb \@tempa \fi \ifx
  \@tempb \@empty \def\@tempb {arXiv}\fi \@ifundefined
  {mn@eprint@\@tempb}{\@tempb:\@tempc}{\expandafter \expandafter \csname
  mn@eprint@\@tempb\endcsname \expandafter{\@tempc}}}

\bibitem[\protect\citeauthoryear{{Baldwin}, {Watkins}, {van der Marel},
  {Bianchini}, {Bellini}  \& {Anderson}}{{Baldwin} et~al.}{sub}]{Baldwin2016}
{Baldwin} A.,  {Watkins} L.~L.,  {van der Marel} R.~P.,  {Bianchini} P.,
  {Bellini} A.,   {Anderson} J.,  sub.

\bibitem[\protect\citeauthoryear{{Baumgardt} \& {Makino}}{{Baumgardt} \&
  {Makino}}{2003}]{BaumgardtMakino2003}
{Baumgardt} H.,  {Makino} J.,  2003, \mn@doi [\mnras]
  {10.1046/j.1365-8711.2003.06286.x}, \href
  {http://adsabs.harvard.edu/abs/2003MNRAS.340..227B} {340, 227}

\bibitem[\protect\citeauthoryear{{Bellini} et~al.,}{{Bellini}
  et~al.}{2014}]{Bellini2014}
{Bellini} A.,  et~al., 2014, \mn@doi [\apj] {10.1088/0004-637X/797/2/115},
  \href {http://adsabs.harvard.edu/abs/2014ApJ...797..115B} {797, 115}

\bibitem[\protect\citeauthoryear{{Bianchini}, {Norris}, {van de Ven},
  {Schinnerer}, {Bellini}, {van der Marel}, {Watkins}  \& J.}{{Bianchini}
  et~al.}{sub}]{Bianchini2016}
{Bianchini} P.,  {Norris} M.~A.,  {van de Ven} G.,  {Schinnerer} E.,  {Bellini}
  A.,  {van der Marel} R.~P.,  {Watkins} L.~L.,   J. A.,  sub.

\bibitem[\protect\citeauthoryear{{Binney} \& {Tremaine}}{{Binney} \&
  {Tremaine}}{2008}]{BinneyTremaine2008}
{Binney} J.,  {Tremaine} S.,  2008, {Galactic Dynamics: Second Edition}.
Princeton University Press

\bibitem[\protect\citeauthoryear{{Da Costa} \& {Freeman}}{{Da Costa} \&
  {Freeman}}{1976}]{DaCostaFreeman1976}
{Da Costa} G.~S.,  {Freeman} K.~C.,  1976, \mn@doi [\apj] {10.1086/154363},
  \href {http://adsabs.harvard.edu/abs/1976ApJ...206..128D} {206, 128}

\bibitem[\protect\citeauthoryear{{Djorgovski}}{{Djorgovski}}{1993}]{Djorgovski1993}
{Djorgovski} S.,  1993, in {Djorgovski} S.~G.,  {Meylan} G.,  eds,
  Astronomical Society of the Pacific Conference Series Vol. 50, Structure and
  Dynamics of Globular Clusters. p.~373

\bibitem[\protect\citeauthoryear{{Downing}, {Benacquista}, {Giersz}  \&
  {Spurzem}}{{Downing} et~al.}{2010}]{Downing2010}
{Downing} J.~M.~B.,  {Benacquista} M.~J.,  {Giersz} M.,   {Spurzem} R.,  2010,
  \mn@doi [\mnras] {10.1111/j.1365-2966.2010.17040.x}, \href
  {http://adsabs.harvard.edu/abs/2010MNRAS.407.1946D} {407, 1946}

\bibitem[\protect\citeauthoryear{{Ferraro} et~al.,}{{Ferraro}
  et~al.}{2012}]{Ferraro2012}
{Ferraro} F.~R.,  et~al., 2012, \mn@doi [\nat] {10.1038/nature11686}, \href
  {http://adsabs.harvard.edu/abs/2012Natur.492..393F} {492, 393}

\bibitem[\protect\citeauthoryear{{Foreman-Mackey}, {Hogg}, {Lang}  \&
  {Goodman}}{{Foreman-Mackey} et~al.}{2013}]{Foreman-Mackey2013}
{Foreman-Mackey} D.,  {Hogg} D.~W.,  {Lang} D.,   {Goodman} J.,  2013, \mn@doi
  [\pasp] {10.1086/670067}, \href
  {http://adsabs.harvard.edu/abs/2013PASP..125..306F} {125, 306}

\bibitem[\protect\citeauthoryear{{Gieles} \& {Zocchi}}{{Gieles} \&
  {Zocchi}}{2015}]{GielesZocchi2015}
{Gieles} M.,  {Zocchi} A.,  2015, \mn@doi [\mnras] {10.1093/mnras/stv1848},
  \href {http://adsabs.harvard.edu/abs/2015MNRAS.454..576G} {454, 576}

\bibitem[\protect\citeauthoryear{{Giersz}}{{Giersz}}{1998}]{Giersz1998}
{Giersz} M.,  1998, \mn@doi [\mnras] {10.1046/j.1365-8711.1998.01734.x}, \href
  {http://adsabs.harvard.edu/abs/1998MNRAS.298.1239G} {298, 1239}

\bibitem[\protect\citeauthoryear{{Gomez-Leyton} \& {Velazquez}}{{Gomez-Leyton}
  \& {Velazquez}}{2014}]{Gomez-LeytonVelazquez2014}
{Gomez-Leyton} Y.~J.,  {Velazquez} L.,  2014, \mn@doi [Journal of Statistical
  Mechanics: Theory and Experiment] {10.1088/1742-5468/2014/04/P04006}, \href
  {http://adsabs.harvard.edu/abs/2014JSMTE..04..006G} {4, 6}

\bibitem[\protect\citeauthoryear{{Gunn} \& {Griffin}}{{Gunn} \&
  {Griffin}}{1979}]{GunnGriffin1979}
{Gunn} J.~E.,  {Griffin} R.~F.,  1979, \mn@doi [\aj] {10.1086/112477}, \href
  {http://adsabs.harvard.edu/abs/1979AJ.....84..752G} {84, 752}

\bibitem[\protect\citeauthoryear{{Harris}}{{Harris}}{2010}]{HarrisCat2010}
{Harris} W.~E.,  2010, arXiv:1012.3224, \href
  {http://adsabs.harvard.edu/abs/2010arXiv1012.3224H} {}

\bibitem[\protect\citeauthoryear{{Heyl}, {Richer}, {Antolini}, {Goldsbury},
  {Kalirai}, {Parada}  \& {Tremblay}}{{Heyl} et~al.}{2015}]{Heyl2015}
{Heyl} J.,  {Richer} H.~B.,  {Antolini} E.,  {Goldsbury} R.,  {Kalirai} J.,
  {Parada} J.,   {Tremblay} P.-E.,  2015, \mn@doi [\apj]
  {10.1088/0004-637X/804/1/53}, \href
  {http://adsabs.harvard.edu/abs/2015ApJ...804...53H} {804, 53}

\bibitem[\protect\citeauthoryear{{Hypki} \& {Giersz}}{{Hypki} \&
  {Giersz}}{2013}]{Hypki2013}
{Hypki} A.,  {Giersz} M.,  2013, \mn@doi [\mnras] {10.1093/mnras/sts415}, \href
  {http://adsabs.harvard.edu/abs/2013MNRAS.429.1221H} {429, 1221}

\bibitem[\protect\citeauthoryear{{Inagaki} \& {Saslaw}}{{Inagaki} \&
  {Saslaw}}{1985}]{InagakiSaslaw1985}
{Inagaki} S.,  {Saslaw} W.~C.,  1985, \mn@doi [\apj] {10.1086/163164}, \href
  {http://adsabs.harvard.edu/abs/1985ApJ...292..339I} {292, 339}

\bibitem[\protect\citeauthoryear{{Inagaki} \& {Wiyanto}}{{Inagaki} \&
  {Wiyanto}}{1984}]{InagakiWiyanto1984}
{Inagaki} S.,  {Wiyanto} P.,  1984, \pasj, \href
  {http://adsabs.harvard.edu/abs/1984PASJ...36..391I} {36, 391}

\bibitem[\protect\citeauthoryear{{Kamann} et~al.,}{{Kamann}
  et~al.}{2016}]{Kamann2016}
{Kamann} S.,  et~al., 2016, preprint, \href
  {http://adsabs.harvard.edu/abs/2016arXiv160201643K} {} (\mn@eprint {arXiv}
  {1602.01643})

\bibitem[\protect\citeauthoryear{{Khalisi}, {Amaro-Seoane}  \&
  {Spurzem}}{{Khalisi} et~al.}{2007}]{Khalisi2007}
{Khalisi} E.,  {Amaro-Seoane} P.,   {Spurzem} R.,  2007, \mn@doi [\mnras]
  {10.1111/j.1365-2966.2006.11184.x}, \href
  {http://adsabs.harvard.edu/abs/2007MNRAS.374..703K} {374, 703}

\bibitem[\protect\citeauthoryear{{King}}{{King}}{1966}]{King1966}
{King} I.~R.,  1966, \mn@doi [\aj] {10.1086/109857}, \href
  {http://adsabs.harvard.edu/abs/1966AJ.....71...64K} {71, 64}

\bibitem[\protect\citeauthoryear{{Kroupa}}{{Kroupa}}{2001}]{Kroupa2001}
{Kroupa} P.,  2001, \mn@doi [\mnras] {10.1046/j.1365-8711.2001.04022.x}, \href
  {http://adsabs.harvard.edu/abs/2001MNRAS.322..231K} {322, 231}

\bibitem[\protect\citeauthoryear{{Lightman} \& {Fall}}{{Lightman} \&
  {Fall}}{1978}]{LightmanFall1978}
{Lightman} A.~P.,  {Fall} S.~M.,  1978, \mn@doi [\apj] {10.1086/156058}, \href
  {http://adsabs.harvard.edu/abs/1978ApJ...221..567L} {221, 567}

\bibitem[\protect\citeauthoryear{{Merritt}}{{Merritt}}{1981}]{Merritt1981}
{Merritt} D.,  1981, \mn@doi [\aj] {10.1086/112891}, \href
  {http://adsabs.harvard.edu/abs/1981AJ.....86..318M} {86, 318}

\bibitem[\protect\citeauthoryear{{Meylan}}{{Meylan}}{1987}]{Meylan1987}
{Meylan} G.,  1987, \aap, \href
  {http://adsabs.harvard.edu/abs/1987A%26A...184..144M} {184, 144}

\bibitem[\protect\citeauthoryear{{Meylan} \& {Heggie}}{{Meylan} \&
  {Heggie}}{1997}]{MeylanHeggie1997}
{Meylan} G.,  {Heggie} D.~C.,  1997, \mn@doi [\aapr] {10.1007/s001590050008},
  \href {http://adsabs.harvard.edu/abs/1997A%26ARv...8....1M} {8, 1}

\bibitem[\protect\citeauthoryear{{Miocchi}}{{Miocchi}}{2006}]{Miocchi2006}
{Miocchi} P.,  2006, \mn@doi [\mnras] {10.1111/j.1365-2966.2005.09842.x}, \href
  {http://adsabs.harvard.edu/abs/2006MNRAS.366..227M} {366, 227}

\bibitem[\protect\citeauthoryear{{Plummer}}{{Plummer}}{1911}]{Plummer1911}
{Plummer} H.~C.,  1911, \mnras, \href
  {http://adsabs.harvard.edu/abs/1911MNRAS..71..460P} {71, 460}

\bibitem[\protect\citeauthoryear{{Shanahan} \& {Gieles}}{{Shanahan} \&
  {Gieles}}{2015}]{ShanahanGieles2015}
{Shanahan} R.~L.,  {Gieles} M.,  2015, \mn@doi [\mnras]
  {10.1093/mnrasl/slu205}, \href
  {http://adsabs.harvard.edu/abs/2015MNRAS.448L..94S} {448, L94}

\bibitem[\protect\citeauthoryear{{Sollima}, {Bellazzini}  \& {Lee}}{{Sollima}
  et~al.}{2012}]{Sollima2012}
{Sollima} A.,  {Bellazzini} M.,   {Lee} J.-W.,  2012, \mn@doi [\apj]
  {10.1088/0004-637X/755/2/156}, \href
  {http://adsabs.harvard.edu/abs/2012ApJ...755..156S} {755, 156}

\bibitem[\protect\citeauthoryear{{Sollima}, {Baumgardt}, {Zocchi}, {Balbinot},
  {Gieles}, {H{\'e}nault-Brunet}  \& {Varri}}{{Sollima}
  et~al.}{2015}]{Sollima2015}
{Sollima} A.,  {Baumgardt} H.,  {Zocchi} A.,  {Balbinot} E.,  {Gieles} M.,
  {H{\'e}nault-Brunet} V.,   {Varri} A.~L.,  2015, \mn@doi [\mnras]
  {10.1093/mnras/stv1079}, \href
  {http://adsabs.harvard.edu/abs/2015MNRAS.451.2185S} {451, 2185}

\bibitem[\protect\citeauthoryear{{Spera}, {Mapelli}  \& {Jeffries}}{{Spera}
  et~al.}{sub}]{Spera2016}
{Spera} M.,  {Mapelli} M.,   {Jeffries} R.~D.,  sub.

\bibitem[\protect\citeauthoryear{{Spitzer}}{{Spitzer}}{1969}]{Spitzer1969}
{Spitzer} Jr. L.,  1969, \mn@doi [\apjl] {10.1086/180451}, \href
  {http://adsabs.harvard.edu/abs/1969ApJ...158L.139S} {158, L139}

\bibitem[\protect\citeauthoryear{{Spitzer}}{{Spitzer}}{1987}]{Spitzer1987}
{Spitzer} L.,  1987, {Dynamical evolution of globular clusters}.
Princeton University Press, Princeton

\bibitem[\protect\citeauthoryear{{Trenti} \& {van der Marel}}{{Trenti} \& {van
  der Marel}}{2013}]{TrentivanderMarel2013}
{Trenti} M.,  {van der Marel} R.,  2013, \mn@doi [\mnras]
  {10.1093/mnras/stt1521}, \href
  {http://adsabs.harvard.edu/abs/2013MNRAS.tmp.2161T} {}

\bibitem[\protect\citeauthoryear{{Vishniac}}{{Vishniac}}{1978}]{Vishniac1978}
{Vishniac} E.~T.,  1978, \mn@doi [\apj] {10.1086/156332}, \href
  {http://adsabs.harvard.edu/abs/1978ApJ...223..986V} {223, 986}

\bibitem[\protect\citeauthoryear{{Wang}, {Spurzem}, {Aarseth}, {Giersz},
  {Askar}, {Berczik}, {Naab}  \& {Kouwenhoven}}{{Wang} et~al.}{2016}]{Wang2016}
{Wang} L.,  {Spurzem} R.,  {Aarseth} S.,  {Giersz} M.,  {Askar} A.,  {Berczik}
  P.,  {Naab} T.,   {Kouwenhoven} R.~S.~M.~B.~N.,  2016, \mn@doi [\mnras]
  {10.1093/mnras/stw274}, \href
  {http://adsabs.harvard.edu/abs/2016MNRAS.tmp...74W} {}

\bibitem[\protect\citeauthoryear{{Watkins}, {van der Marel}, {Bellini}  \&
  {Anderson}}{{Watkins} et~al.}{2015a}]{Watkins2015}
{Watkins} L.~L.,  {van der Marel} R.~P.,  {Bellini} A.,   {Anderson} J.,
  2015a, \mn@doi [\apj] {10.1088/0004-637X/803/1/29}, \href
  {http://adsabs.harvard.edu/abs/2015ApJ...803...29W} {803, 29}

\bibitem[\protect\citeauthoryear{{Watkins}, {van der Marel}, {Bellini}  \&
  {Anderson}}{{Watkins} et~al.}{2015b}]{Watkins2015b}
{Watkins} L.~L.,  {van der Marel} R.~P.,  {Bellini} A.,   {Anderson} J.,
  2015b, \mn@doi [\apj] {10.1088/0004-637X/812/2/149}, \href
  {http://adsabs.harvard.edu/abs/2015ApJ...812..149W} {812, 149}

\bibitem[\protect\citeauthoryear{{Wilson}}{{Wilson}}{1975}]{Wilson1975}
{Wilson} C.~P.,  1975, \mn@doi [\aj] {10.1086/111729}, \href
  {http://adsabs.harvard.edu/abs/1975AJ.....80..175W} {80, 175}

\bibitem[\protect\citeauthoryear{{Woolley}}{{Woolley}}{1954}]{Woolley1954}
{Woolley} R.~V.~D.~R.,  1954, \mnras, \href
  {http://adsabs.harvard.edu/abs/1954MNRAS.114..191W} {114, 191}

\bibitem[\protect\citeauthoryear{{Zocchi}, {Bertin}  \& {Varri}}{{Zocchi}
  et~al.}{2012}]{Zocchi2012}
{Zocchi} A.,  {Bertin} G.,   {Varri} A.~L.,  2012, \mn@doi [\aap]
  {10.1051/0004-6361/201117977}, \href
  {http://adsabs.harvard.edu/abs/2012A%26A...539A..65Z} {539, A65}

\makeatother
\end{thebibliography}

\section*{appendix}
 \renewcommand{\theequation}{A\arabic{equation}}
  \setcounter{equation}{0}  

The exponential fitting function proposed in Sect.~3.1 is  physically motivated by two asymptotic behaviours of the central value of the velocity dispersion profile $\sigma(m)$, in the limit of low and high mass, respectively. Such behaviors can be studied in detail by taking advantage of the analytical framework provided by appropriate distribution function-based equilibria. Previous dynamical studies have indeed showed that multi-mass, lowered isothermal models (e.g., \citealp{Woolley1954,King1966,Wilson1975}, and more recently \citealp{Gomez-LeytonVelazquez2014,GielesZocchi2015}, hereafter GZ15) offer a successful description of Galactic globular clusters, even in different relaxation conditions (e.g., see \citealp{DaCostaFreeman1976,GunnGriffin1979,Meylan1987,Sollima2012}). These equilibria are characterized by multiple mass components, which are traditionally defined in terms of a set of relations between the velocity scales $s_j$ and the masses $m_j$ of the different components, such that $m_j\,s_j^2=m_i\,s_i^2$. As previously noted (\citealp{Merritt1981,Miocchi2006}, GZ15), we emphasize that such a prescription does {\it not} enforce a condition of full energy equipartition in the resulting configurations, neither locally nor globally.

This class of models allow us to derive the velocity dispersion profile in closed analytical form (i.e., as a function of the potential), which may be expressed in terms of appropriate special functions. For the reader's convenience, here we will adopt the same notation used by \citet{GielesZocchi2015}, in which the central value of the dimensionless velocity dispersion of the component $j$ is given by: 
\begin{equation}\label{sig}
\hat{\sigma}_{1d\,j\,0} =  \frac{1}{\mu_j^\delta}\frac{E_{\gamma}( g+5/2; \mu_j^{2\delta} \hat{\phi}_0)}{E_{\gamma}( g+3/2; \mu_j^{2\delta} \hat{\phi}_0)},    
\end{equation}
where $\hat{\phi}_0=\hat{\phi}(\hat{r}=0)$ is the depth of the central potential well (i.e., a measure of the central concentration), $\mu_j=m_j/\bar{m}$ is the dimensionless mass of component $j$, and its normalization is given by the central density weighted mean-mass $\bar{m}= \Sigma_j\,m_j\rho_{0j}/\Sigma_j\rho_{0j}$. The function $E_{\gamma}$ is a convenient piecewise definition of the modified lowered incomplete Gamma function, introduced by \citet{Gomez-LeytonVelazquez2014} (see also Eq.~(2) and App.~D1 of GZ15). The parameter $g$ sets the continuity properties of the truncation prescription of the distribution function (see Eq.~(1) of GZ15), in such a way that, in the isotropic limit, $g=0,1,2$ correspond to the usual \citet{Woolley1954}, \citet{King1966}, and non-rotating \citet{Wilson1975} multi-mass models, respectively. Finally, the parameter $\delta$ is defined so that $m_js_j^{1/\delta}=m_i\,s_i^{1/\delta}$; for $\delta=1/2$, such a relation reduces to the condition usually adopted in the literature.

By considering the regime $\mu_j \ll 1$ (i.e., $m_j \ll \bar{m}$), the asymptotic behavior of the function indicated in Eq.~(\ref{sig}) can be easily calculated up to the order $\mathcal{O}(\mu_j^{4 \delta})$ (i.e., second order in $\mu_j$, for $\delta=1/2$):
\begin{align}
\hat{\sigma}_{1d\,j\,0} &\sim \left[ \frac{(g+3/2)\Gamma(g+3/2)}{(g+5/2)\Gamma(g+5/2)}\right]^{1/2}\hat{\phi}_0^{1/2}  \nonumber \\
 & \left[ 1 - \frac{1} {2(g+5/2)(g+7/2)}\, \hat{\phi}_0 \,\mu_j^{2\delta}+ \nonumber \right.\\ 
 & \left. + \,\frac{6+3(g+5/2)-4(g+5/2)^2}{8(g+5/2)^2(g+7/2)^2(g+9/2)}\,\hat{\phi}^2_0 \mu_j^{4\delta}\right]
\end{align}   
where $\Gamma$ denotes the Gamma function. We stress that, in the limit $\mu_j \rightarrow 0$, the central velocity dispersion (for a chosen value of the truncation parameter $g$) tends to a constant value, which depends only on the central concentration $\hat{\phi}_0$. The limiting values for the traditional Woolley, King, and isotropic Wilson models are recovered as $\hat{\sigma}_{1d\,j\,0} \sim A(g)^{1/2} \hat{\phi}_0^{1/2}$, with $A(g)=2/(5+2g)$ for $g=0,1,2$ (to be compared, e.g., with the central values of the models depicted in Fig.~9 of GZ15). Such an asymptotic behavior in the regime of low stellar masses informed our choice for the expression of the fitting function introduced in Eq.~(1), which, for $m \ll m_{eq}$, may be expressed as 
\begin{equation}
\sigma \sim \sigma_0\left[1 -\frac{1}{2} \frac{m}{m_{eq}} + \frac{1}{8}  \left(\frac{m}{m_{eq}}\right)^2\right],
\end{equation}
which is the first terms of a Taylor expansion of $\sigma=\sigma_0\exp(-1/2\,m/m_\mathrm{eq})$ 

Similarly, by considering the regime $\mu_j \gg 1$ (i.e., $m_j \gg \bar{m}$), the asymptotic behavior of the function indicated in Eq.~(\ref{sig}) is simply given by $\hat{\sigma}_{1d\,j\,0} \sim 1/\mu_j^{\delta}$. For the typical value $\delta=1/2$, this corresponds to the traditional scaling $\hat{\sigma}_{1d\,j\,0} \sim m_j^{-1/2}$ (see also Sect.~3.2.1 of GZ15), with a coefficient, $\bar{m}^{1/2}$, which is, once again, a function of the central concentration alone (for a chosen value of the parameter $g$). Such a behavior motivates the piecewise definition of our fitting function in the regime of higher masses (i.e., $m \gg m_{eq}$).

\end{document}